# Human-AI collaborative autonomous synthesis with pulsed laser deposition for remote epitaxy

*Asraful Haque[1], Daniel T. Yimam[1], Jawad Chowdhury[1], Ralph Bulanadi[1], Ivan Vlassiouk[1], John Lasseter[1], Sujoy Ghosh[1], Christopher M. Rouleau[1], Kai Xiao[1], Yongtao Liu[1], Eva Zarkadoula[1], Rama K. Vasudevan[1], Sumner B. Harris[1]\**

1. Center for Nanophase Materials Sciences, Oak Ridge National Laboratory, Oak Ridge, Tennessee 37831, United States.

   *Correspondence should be addressed to: harrissb@ornl.gov

Notice: This manuscript has been authored by UT-Battelle, LLC, under Contract No. DE-AC05- 00OR22725 with the U.S. Department of Energy. The United States Government retains and the publisher, by accepting the article for publication, acknowledges that the United States Government retains a non-exclusive, paid-up, irrevocable, world-wide license to publish or reproduce the published form of this manuscript, or allow others to do so, for United States Government purposes. The Department of Energy will provide public access to these results of federally sponsored research in accordance with the DOE Public Access Plan ( http://energy.gov/downloads/doe-public-access-plan ).

## ABSTRACT

Autonomous laboratories typically rely on data-driven decision-making, occasionally with human-in-the-loop oversight to inject domain expertise[1-6]. Fully leveraging AI agents, however, requires tightly coupled, collaborative workflows spanning hypothesis generation, experimental planning, execution, and interpretation. To address this, we develop and deploy a human-AI collaborative (HAIC) workflow that integrates large language models for hypothesis generation and analysis, with collaborative policy updates driving autonomous pulsed laser deposition (PLD) experiments for remote epitaxy of $BaTiO_3$/graphene. HAIC accelerated the hypothesis formation and experimental design and efficiently mapped the growth space to graphene-damage. *In situ* Raman spectroscopy reveals that chemistry drives degradation while the highest energy plume components seed defects, identifying a low-$O_2$ pressure low-temperature synthesis window that preserves graphene but is incompatible with optimal $BaTiO_3$ growth. Thus, we show a two-step Ar/$O_2$ deposition is required to exfoliate ferroelectric $BaTiO_3$ while maintaining a monolayer graphene interlayer. HAIC stages human insight with AI reasoning between autonomous batches to drive rapid scientific progress, providing an evolution to many existing human-in-the-loop autonomous workflows.

**INTRODUCTION**

Autonomous and robotic laboratories are transforming materials synthesis by combining high-throughput experimentation and computation with machine learning to optimize processes[1,2]. The prevailing paradigm imbues automated synthesis and characterization laboratories with decision-making driven by artificial intelligence (AI) to realize desired molecular and material properties while minimizing costly human intervention[3]. However, human-in-the-loop (HITL) workflows have been shown to increase efficiency in industry[4] and in materials characterization, such as X-ray phase mapping[5] and atomic force microscopy (AFM)[6], by bounding or initializing optimization procedures with human intuition within an automated experimental loop[7]. This improvement can be rationalized as arising from the incorporation of expert prior knowledge rather than the typically uninformative priors assumed by standard optimization algorithms. However, existing HITL workflows still suffer from significant drawbacks, with most being limited to an interventionist approach in which humans either seed initial experiments, modify parameter spaces, or inject knowledge at predefined points, limiting their utility. To leverage the rapidly developing capabilities of modern AI agents, more tightly coupled, collaborative workflows are required across all stages of the experimental process, from hypothesis generation and experimental planning to results interpretation that go beyond existing HITL workflows.

Here, we evolve HITL into human-AI collaborative (HAIC) workflows that couple human expertise, large language models (LLMs), and autonomous systems through mixed-initiative, between-batch loops. This approach is amplified by using LLMs alongside autonomous systems as "co-scientists"[8,9] to help generate hypotheses, plan experiments[10], and analyze data, especially when retrieval-augmented generation[11] (RAG) grounds the model in relevant scientific corpora[12]. This flexibility is particularly suited to thin-film synthesis, where large parameter spaces with sparse prior data make defining robust success metrics challenging, especially when exploring new applications or materials systems.

Notably, thin-film synthesis poses a different set of challenges than materials discovery and requires a tightly coupled collaborative workflow. Success metrics may require complex measurements with bespoke analysis, defined iteratively rather than *a priori*, and are subject to significant data complexity (noise, unknowns, intangibles). Further, general mechanistic and experimental databases for thin-film synthesis are lacking, and literature-based recipes are often too unreliable to predict performance metrics[13,14], making the incorporation of prior knowledge difficult. As data accumulates, the workflow must evolve to accommodate process improvement[15] in all but the simplest scenarios to be effective. Addressing these challenges is essential to broaden the success and applicability of autonomous platforms, especially for thin-film synthesis.

Our work applies HAIC to autonomous synthesis, addressing key challenges in the emerging field of remote epitaxy (RE) of complex oxides by pulsed laser deposition (PLD). In RE, a two-dimensional interlayer, such as graphene, is placed on a single-crystal substrate, enabling the epitaxial growth of single-crystalline films that can be exfoliated as thin membranes and integrated on arbitrary substrates[16]. However, PLD of complex oxides such as $BaTiO_3$ (BTO) requires growth conditions that destroy monolayer graphene interlayers, so successful PLD RE has relied on bilayer graphene to retain film transferability[17,18]. This poses a serious challenge for RE in PLD (and molecular beam epitaxy) because the transmitted electrostatic potential of the substrate through graphene is small (~10-20 meV) and short-ranged (~2 Å)[19]. Consequently, >1 graphene layers weaken or eliminate remote film alignment, yielding poorly oriented or polycrystalline films[20], depending on the substrate's ionicity/polarity[21,22]. Comprehensive synthesis studies are essential to understand the limitations of oxide RE with PLD and aid the development of alternative strategies to enable remote epitaxy for arbitrary materials.

We deploy a HAIC strategy by combining state-of-the-art LLMs with an autonomous PLD system equipped with *in situ* diagnostics to understand how complex-oxide growth conditions drive graphene damage during RE, using BTO as a test case. LLM-assisted hypothesis generation, experimental design, and iterative process refinements enabled the autonomous campaign to efficiently map the parameter space and identify the BTO growth regime that minimizes graphene damage. Targeted *in situ* Raman experiments further elucidate the interplay of chemical interactions and ballistic damage in graphene defect generation during BTO deposition, providing a mechanistic basis for $O_2$-mediated RE with PLD. We also show that a two-step growth sequence minimizes graphene damage while yielding ferroelectric BTO. Together, these results show that a HAIC approach can rapidly close knowledge gaps in new fields and demonstrate its potential for application in any autonomous process that benefits from flexible, iterative policy updates led collaboratively by humans and AI.

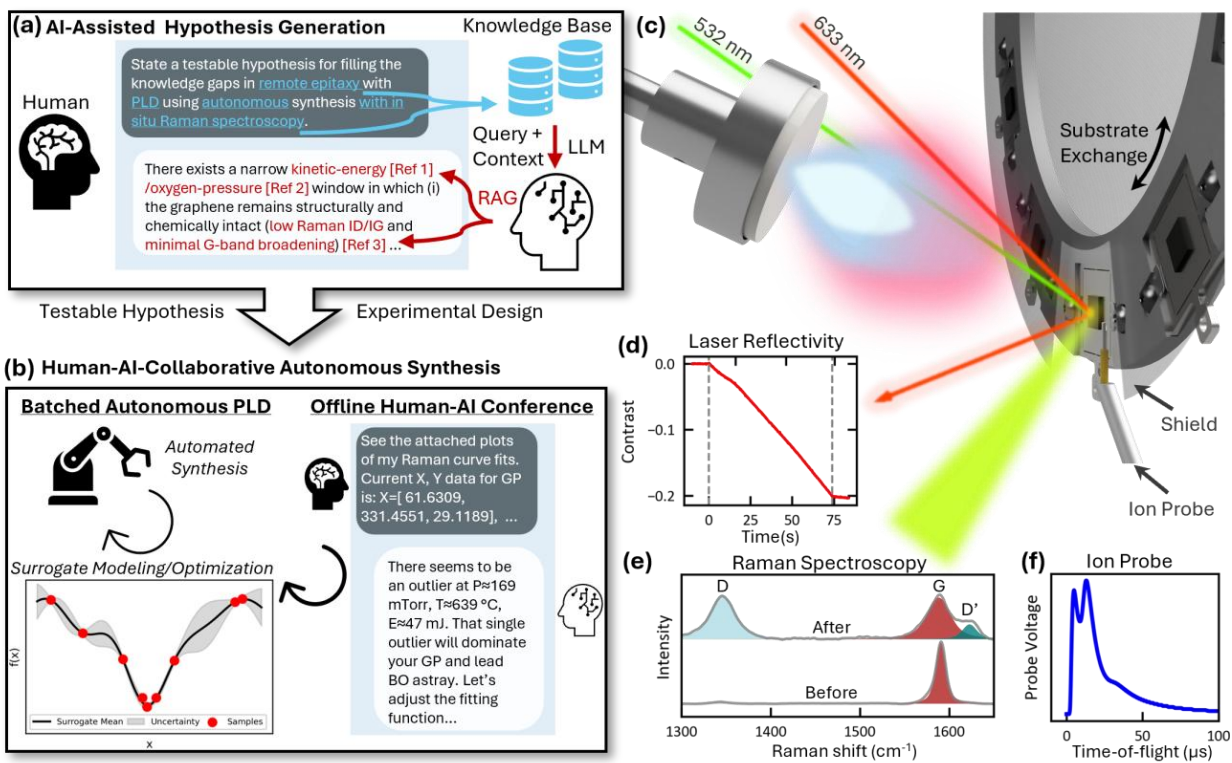

**Figure 1**. AI-Assisted hypothesis generation and human-AI collaborative (HAIC) autonomous synthesis. (a) A curated domain-specific knowledge base is queried via retrieval-augmented generation (RAG), and the human expert and LLM engage in an iterative refinement loop to converge on a final testable hypothesis and initial experimental design. (b) HAIC autonomous synthesis is conducted which combines fully autonomous synthesis with human-AI conferences to cooperatively analyze data and refine the active learning procedure. In online active learning, an autonomous batch of PLD synthesis with *in situ* characterization is conducted using Bayesian optimization to drive the sequence of experiments in the batch. The offline conferences between the human and LLM are conducted to review the collected data suggest and adjust the surrogate model/acquisition function/target as needed back into the next batch of autonomous PLD.

## RESULTS AND DISCUSSION

**Human-AI Collaborative Autonomous Synthesis Workflow**

The HAIC autonomous synthesis workflow consists of 3 parts: AI-assisted generation of a testable hypothesis and a proposed experiment; fully autonomous PLD synthesis guided by Bayesian optimization (BO); and human-AI conferences for iterative data analysis and process improvement. First, we employed an LLM with a RAG pipeline to assist in literature summarization for hypothesis generation by prompting the LLM to identify research gaps in RE with PLD and to propose an experiment based on our experimental capabilities. **Figure 1a** shows a schematic of this AI-assisted hypothesis generation procedure. We compiled a curated knowledge base (KB) spanning scientific domains related to: RE in general, RE with PLD, Raman spectroscopy of graphene for defect analysis, PLD in general, PLD of BTO, autonomous experimentation, and Bayesian optimization. During RAG, an embedding-based similarity search is used to retrieve KB passages from a diversity of sources that have high similarity to the query. The retrieved context is appended to the prompt to condition the LLM to prioritize information from the curated KB before defaulting to built-in knowledge or web search tools.

After hypothesis generation, we iteratively synthesized batches of samples via autonomous PLD, whose growth conditions were sequentially selected by BO, and reviewed the results via offline conferences with the LLM by prompting it with text, plot images, and raw data (**Figure 1b**). Here, "offline" specifies that the data transfer and prompts to the LLM were done by a human and not directly automated by the PLD software. Directly calling LLM-based analysis via the PLD software is simple to implement (an "agentic AI" platform), but the risk of costly, inevitable faults caused by the LLM in high-stakes experiments is unacceptable. In this scenario, it is best to keep LLMs as "offline advisors" until their capabilities improve. Agentic platforms for lower-risk experiments like AFM could be immediately empowering. However, the true benefit of this step is in the human-AI interaction via unstructured "chats", collaborative data analysis, and AI-assisted human decision-making.

A schematic of the autonomous PLD system is shown in **Figure 1c** with details found elsewhere[23]. This system combines automated PLD synthesis, multimodal *in situ* diagnostics, and substrate exchange with online machine learning for on-the-fly decision making post-growth or in real-time[24]. Critically, the system is equipped with *in situ* Raman spectroscopy that is used to evaluate how the monolayer graphene interlayer accumulates damage during processing and/or BTO deposition (**Figure 1e**). The film growth rate varies by orders of magnitude depending on the conditions selected by BO. We use *in situ* laser reflectivity to control the thickness of each film to be nominally identical for fair comparisons across the dataset (**Figure 1d**). Lastly, an ion probe next to the substrate provides a measure of the kinetic energy of the plume, which is known to be a major factor in doping and damage of 2D materials in PLD[25-27] (**Figure 1f**). Together, these diagnostics coupled with autonomy provide a unique opportunity to explore the relationship between oxide RE conditions and interfacial graphene damage.

The final AI-assisted hypothesis we selected is summarized here (paraphrased, **Note S1**). There exists a well-defined window of PLD growth parameters, specifically $O_2$ background pressure ($pO_2$), laser fluence (F), and substrate temperature ($T_s$), in which (a) the kinetic energy of the plume species falls below the displacement threshold of graphene (<~22 eV/C-atom[28]) and (b) $BaTiO_3$ still nucleates in registry to the substrate through the graphene interlayer. In that window, the Raman defect metric $I_D/I_G$, the intensity ratio of graphene's D and G bands measured by Raman spectroscopy, will remain small, and *ex situ* structural probes (i.e., XRD) will confirm remote epitaxy. Outside this window, either graphene damage increases or film registry is lost. The LLM suggests that fast neutral species[29] in the PLD plume will induce graphene defects (low $pO_2$, high F), high $T_s$ (> 800 °C) will oxidize the graphene[18], and too-low $T_s$ will fail to crystallize BTO. To test this hypothesis, the LLM-suggested autonomous experiment was to deposit ~10 nm of BTO onto monolayer graphene/$SrTiO_3$ substrates with various $pO_2$ (50-250 mTorr), F (1.2-2.2 J cm$^{-2}$), and $T_s$ (650-800 °C) and conduct BO to minimize $I_D/I_G$.

We modified the AI-suggested experiment to explore lower $T_s$ = 300-800 °C and used $SiO_2$/Si substrates to achieve high optical contrast for *in situ* laser reflectivity and to mitigate the difficulty in reliable quantification of $I_D/I_G$ on $SrTiO_3$ (STO). The Raman R-band of STO dominates in the same region as the graphene D/G bands[18]. $SiO_2$/Si does not provide the driving force for epitaxy, but we still found that BTO can grow oriented in some situations (**Figure S1**). We supplemented the autonomous experiments with additional experiments on STO to test for substrate registry under replicated conditions. Finally, $I_D/I_G$ is not an ideal optimization metric because: it cannot distinguish between defective graphene and amorphous carbon[30]; nor does it account for graphene loss via etching/oxidation; and does not account for variability in the initial graphene quality. Instead, we define a quality score based on the change in $I_D/I_G$, $\Delta = (I_D/I_G)_{final} - (I_D/I_G)_{initial}$ and define a retention ratio $r_G = I_{G,final}/I_{G,initial}$ which approximates how much graphene remains after deposition. The graphene quality is defined as $score = r_G/(1+\Delta)$. We use BO to maximize this score, which is bounded from 0 to 1, where score = 1 indicates that the graphene was unchanged by the BTO deposition and score = 0 indicates that the graphene was destroyed.

**Autonomous synthesis of BaTiO₃ with a monolayer graphene interlayer**

**Figure 2** shows the results of the human-AI-collaborative autonomous synthesis campaign of BTO on monolayer graphene/$SiO_2$/Si substrates. We grew a total of 37 films across 6 batches (phases) of 3-10 samples with no repeated growth conditions. Offline human-AI conferences were conducted after each phase. The first 3 sample conditions were chosen at random, and then we began BO using a Gaussian process (GP) surrogate model and acquisition function to propose the next (single) growth condition to maximize the graphene quality score. The final GP surrogate predictions, projected into each 2D parameter plane, are shown in **Figure 2a-c** (**Figure S2** shows kernel-free surrogate model fits). The results clearly show that retaining a high-quality monolayer graphene interlayer in an O$_2$ environment requires low $pO_2$ and low $T_s$ and is independent of F over the explored range. After phase I, the offline human-AI conference resulted in sequential refinement of the graphene Raman spectrum fitting procedure and provided monitoring of the optimization progress, identifying an early outlier result that strongly skewed the GP (**Note S2**).

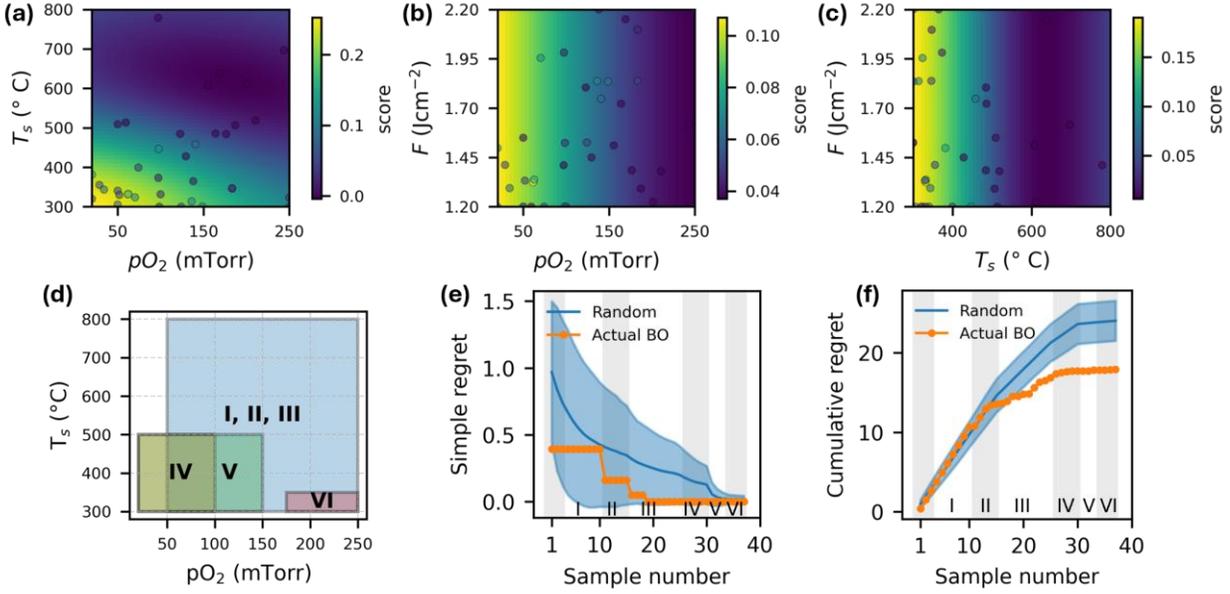

**Figure 2**. Human-AI collaborative (HAIC) autonomous synthesis campaign of BaTiO$_3$ on monolayer graphene/SiO$_2$/Si substrates. The final Gaussian process (GP) surrogate model predictions for the graphene quality $score = r_G/(1 + \Delta)$ ($r_G$ is the retention ratio and $\Delta$ is change in $I_D/I_G$) in the parameter space are shown projected into the (a) substrate temperature $T_s$ vs. oxygen pressure $pO_2$ plane, the (b) laser fluence F vs. $pO_2$ plane, and the (c) F vs. $T_s$ plane. To retain high monolayer graphene quality, low-$T_s$, low-$pO_2$ growth is required, and is essentially independent of F. The HAIC synthesis campaign used offline human-AI conferences resulting in experimental phases with sequential improvements to Raman analysis and dynamic parameter bounds for batches of samples. The parameter bounds for optimization in each of the 6 experiment phases are shown in (d). The simple (e) and cumulative (f) regret of the actual Bayesian optimization (BO) results compared to 500 random trials (using the final GP surrogate model as ground truth) confirms the BO outperformed random sampling, with each phase represented by the gray/white shaded regions. The 95% confidence bound is plotted with the mean of the random trials.

In parallel, human intuition was used to adjust the parameter bounds to align the BO exploration toward regions of interest, which are shown per phase in **Figure 2d**.

After the experiment, the final GP model was considered the "ground truth" and BO performance was quantified by comparing to 500 random sampling trials. In **Figure 2e,f**, the random curves are the mean of 500 trials, and the shaded area is the 95% confidence interval. We analyze regret at evaluation $t$: $r_t = f_p^* - f_p(\mathbf{x_t})$ where $f_p^*$ is the GP optimum in phase $p$ within the active bounds and $f_p(x_t)$ are the GP predictions at that same step/phase/space. We report simple regret $s_t = min_{i \leq t}(r_i)$ and cumulative regret $R_t = \sum_{i=1}^{t} r_i$. Small steps in simple regret occur at the boundaries between phases, a result from updated training data from refined curve fitting as well as GP surrogate improvement as data volume increases. **Figure 2e** shows the simple regret drops sharply after phase 1 and is near 0 after about 15 samples, beating the mean of the random trial baseline. **Figure 2f** shows the cumulative regret flattens at the same time, whereas the random case continues to increase. On average, BO reaches the optimum faster and accumulates fewer samples with low scores. Together, the surrogate maps and regret analysis demonstrate that the HAIC BO strategy efficiently discovered the low-$pO_2$, low-$T_s$ growth window while reducing frivolous experiments relative to random sampling.

We found the primary benefit of the human-AI conferences was the rapid iterative refinement of curve fitting methods used to calculate the Raman score (**Note S3, Figure S3, Figure S4**). A major challenge in autonomous experiments is determining a "scalarizer" to convert high-dimensional data into a scalar value for optimization[7]. Here, the Raman score required a multi-peak curve fitting scalarizer. Generally, the data needed to develop a fitting routine robust against "unknown knowns" and "unknown unknowns" present in real spectra is not available prior to conducting experiments. This development may be needlessly time consuming if attempting to account for all conceivable data complexities. A better approach is to begin collecting data and refine on-the-fly with a HAIC workflow. Without the offline human-AI conferences, the experiment (as initialized) would have failed if run in a fully autonomous modality due to early outliers caused by a non-robust scalarizer.

**Contributions of plume kinetic energy and reactive chemistry to graphene damage**

We probe the origin of the graphene damage with *in situ* Raman spectroscopy by correlating defect metrics with plume kinetic energy, background gas, and temperatures, and we support this interpretation with molecular dynamics (MD) simulations. **Figure 3a** compares ion probe traces in 60 mTorr $O_2$/Ar and vacuum. The probe traces exhibit a clear plume splitting into "fast" and "slow" components[31]. In vacuum, the plume has a "fast" component (~108.5 eV/Ba-atom, most probable kinetic energy (MPKE) taken from the time-of-flight of the peak) arising from ballistic species with no gas-phase scattering. In Ar and $O_2$, a fraction of the "fast" component is scattered, producing an additional "slow" component (~6.6 eV/Ba-atom MPKE in both Ar and $O_2$). The "slow" component MPKE is tunable with $pO_2$ and F but is always below the C displacement threshold (**Figure S5**). **Figure 3b** shows the change in $I_D/I_G$ during BTO deposition in 60 mTorr $O_2$/Ar at temperatures from 20-500 °C. The $I_D/I_G$ rises steeply in $O_2$, especially at elevated temperatures, while Ar produces a markedly shallower slope. The plume kinetic energy distribution is similar in both gases at this pressure, indicating that an Ar background suppresses the degradation that occurs rapidly in $O_2$. Bombardment of the graphene/$SiO_2$ by the "fast" component was simulated by MD, which shows that some vacancies and disorder are caused by the plume, but it does not destroy the graphene (**Figure 3c**). We also tested the maximum energy (plume leading edge, 318 eV/Ba-atom) which caused more disorder (**Figure S6**) but without complete destruction.

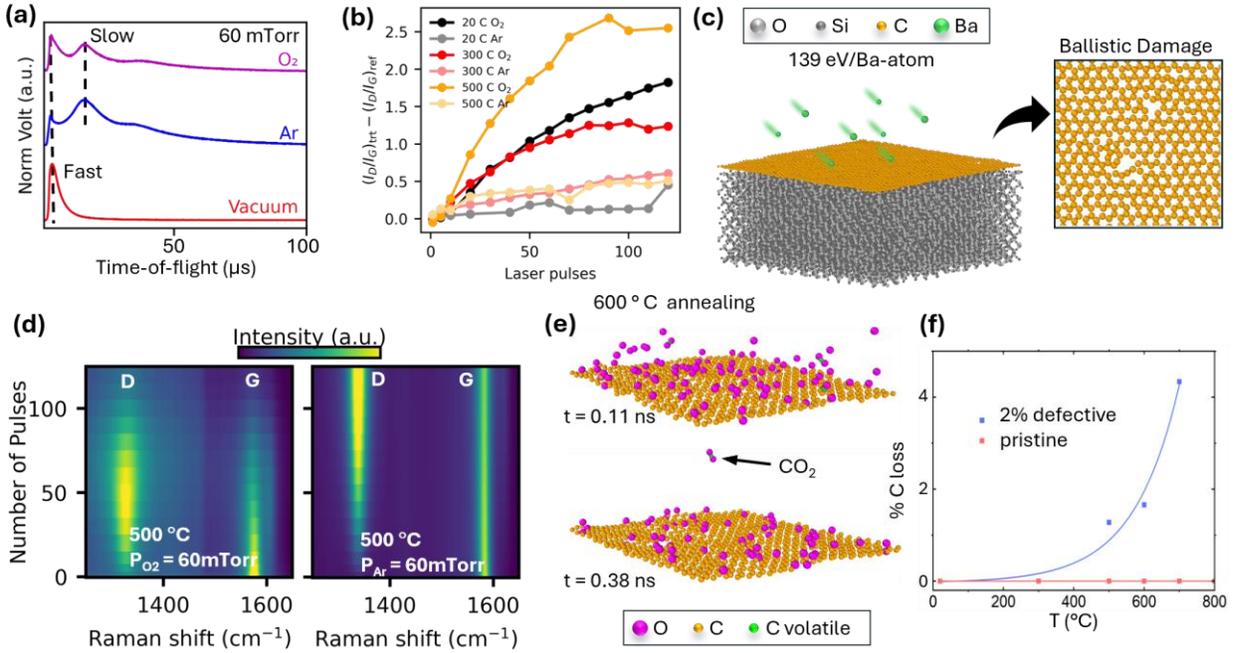

**Figure 3.** The interplay of chemical reactivity and plume-induced ballistic damage dictate graphene degradation. (a) Ion probe traces (4.5 cm from the target) in $O_2$, Ar, and vacuum. (b) Evolution of the Raman defect metric $\Delta$ (the change in $I_D/I_G$) versus laser pulses in $O_2$ and Ar at different substrate temperatures (60 mTorr). (c) Molecular dynamics (MD) simulations of Ba bombardment on graphene/$SiO_2$ shows C vacancy generation and disorder. (d) Raman heatmaps vs laser pulses at 500 °C for $O_2$ and Ar show that graphene is etched during $BaTiO_3$ deposition in $O_2$ but not in Ar. (e) MD simulations of graphene oxidation shows that 2% C-vacancy graphene is rapidly etched at high temperatures while pristine graphene is unreactive. (f) C loss vs temperature calculated from MD simulations show a thermally activated oxidation process.

The ballistic damage alone does not account for the complete etching of graphene that we observe at high temperatures. **Figure 3d** shows the evolution of the Raman spectrum vs pulses during BTO deposition at 500 °C in $O_2$ and Ar. In $O_2$, the graphene D mode intensity rapidly grows, decreases, and vanishes concomitantly with the G mode, indicating a full amorphization trajectory and etching of graphene. In Ar, defects form but the graphene is not amorphized/etched. At 300 °C and 20 °C, we do not observe etching, suggesting a thermally activated oxidation process (**Figure S7**). We also ran no-deposition controls (**Figure S8**). During annealing in Ar and $O_2$ (**Figure S8a, b**), no D band appears up to 700 °C, indicating thermal cycling alone does not create significant point defects.

MD simulations of the oxidation of pristine and 2% C-vacancy graphene show that O species adsorb to graphene and attack C-vacancy sites, forming $CO_2$ that desorbs over time (**Figure 3e**). We conducted temperature-dependent simulations from 20-800 °C, measured the number of C atoms that leave the system, and found agreement with experiments (**Figure 3f**). First, MD confirms that pristine graphene does not oxidize up to 800 °C. Second, we observe a thermally activated oxidation process in which the number of C atoms leaving the monolayer increases exponentially with temperature, with complete etching at 800 °C.

These comparisons make the kinetic energy-chemistry-damage relationship explicit and underscore that the damage induced by plume kinetic energy is independent of the damage via chemical reactivity, though it must assist oxidation in an $O_2$ environment. Energetic O species in the plume are not a primary cause of graphene oxidation; only the highest MPKE plume components cause ballistic damage to the graphene. The fast component of the plume must still damage the graphene at higher pressures, but the rate is significantly decreased due to order-of-magnitude attenuation in the number of fast particles. However, these plume-induced vacancies are highly reactive to $O_2$, resulting in a defect cascade that accelerates with temperature.

**Two-step deposition for epitaxial ferroelectric BTO with monolayer graphene interlayer**

**Figure 2** established that preserving monolayer graphene during oxide PLD requires low-$pO_2$ and low-$T_s$, while **Figure 3** traced the protection afforded by Ar at least up to 700 °C. Building on this, we first assessed whether one-step $O_2$ growth at reduced $T_s$ could yield crystalline BTO on (001) STO. XRD θ-2θ scans in **Figure 4a** show that the BTO (002) reflection is weak or absent below 500 °C and strengthens as $T_s$ increases, indicating that the minimum temperature required for BTO crystallization on STO is between 400-500°C. This constraint conflicts with preservation of graphene in $O_2$, motivating a two-step approach in which a seed-layer is first grown in Ar and followed by continued growth in $O_2$, suggested by Lee et al[32]. The results of this two-step growth at 700 °C are shown in **Figure 4b (Figure S9a** shows X-ray reflectivity). The BTO (002) peak confirms that the two-step route preserves BTO crystallinity, even though it was partially grown in an O-deficient environment. Raman measurements verify that graphene is not fully etched by this process: the Gr/$SiO_2$/Si reference sample exhibits a strong G mode with low D to start, and after two-step BTO growth at 700 °C the G band remains visible with a moderate D band (**Figure 4c-d**), consistent with partial disorder. This approach also enabled exfoliation of the BTO film using monolayer CVD-grown, wet-transferred graphene, as confirmed by the XRD and optical image in Figure 4b (literature comparison in Table S2).

Finally, piezoresponse force microscopy (PFM) utilizing interferometric detection to minimize cantilever-driven electrostatics show that BTO films grown by the two-step deposition method maintain a ferroelectric polarization. Application of -10 V through the scanning probe tip reverses the polarization, suggesting the sample polarization is initially down-oriented (application of +10 V shows no change from the non-switched regions) (**Figure 4d, S9b**). Ferroelectricity is further established through switching spectroscopy PFM (SS-PFM) measurements, which reveal hysteresis behavior with 180° phase reversal and a coercive voltage of ~ ± 1.5 V (**Figure 4e**); this behavior disappears when heated above the Curie temperature (**Figure 4f**). Together with the *in situ* Raman study (**Figure 3**), these results show that Ar in the first step limits both ballistic damage and oxidation of graphene, and the subsequent $O_2$ step maintains O-stoichiometry to achieve ferroelectric BTO.

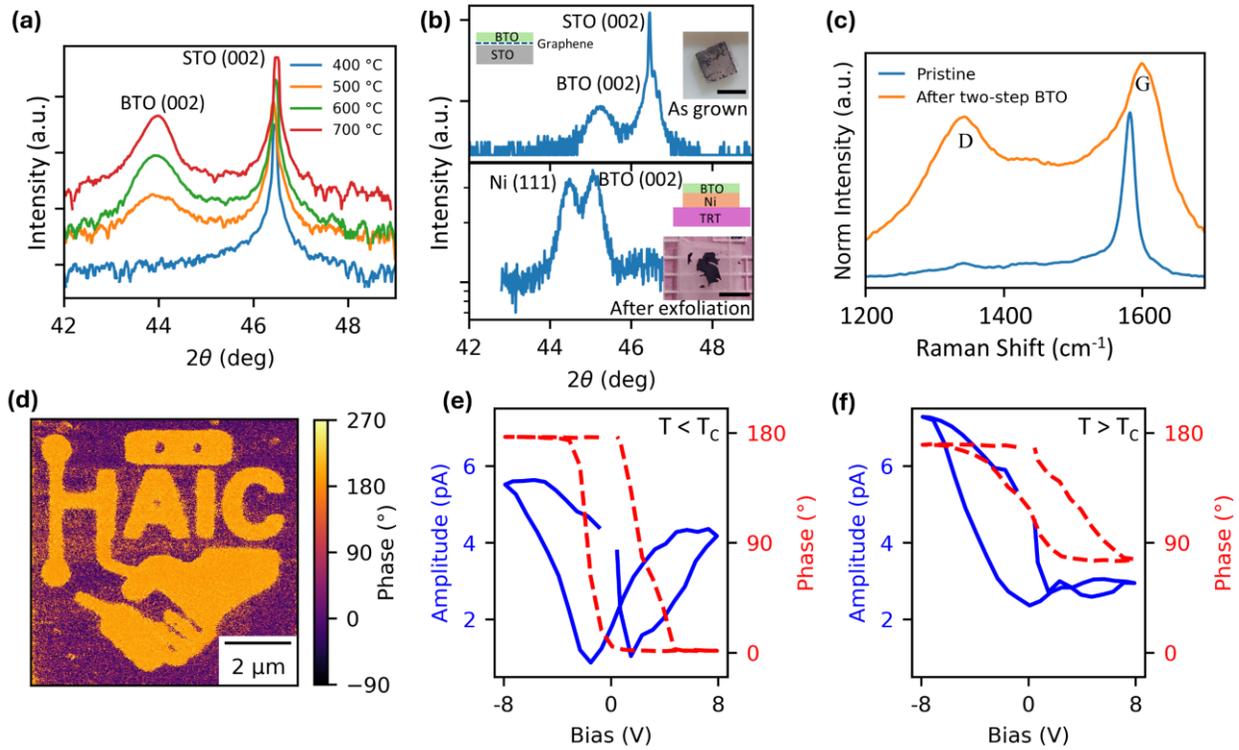

**Figure 4:** Two step growth enables crystalline BaTiO3 while preserving graphene. X-ray diffraction θ-2θ scans around SrTiO3 (002) (a) one-step BaTiO3 growth on SrTiO3 with different substrate temperature, (b) 50 nm thick BTO film by two-step growth (first in Ar, then in O2) at 700 °C followed by exfoliation of the BTO/Ni (stressor layer) exfoliation with the thermal release tape (TRT). Inset shows the cross section of the heterostructure and optical image (scale bar 5mm) (c) Raman of Gr/SiO2/Si reference at 150 °C before deposition, and after two-step growth shows a retained G band and moderate D band increase, indicating graphene is not fully etched. (d) Patterning via piezoresponse force microscopy (PFM) shows clear polarization reversal. (e) Switching spectroscopy PFM (SS-PFM) amplitude and phase taken below the Curie temperature ($T_C$) show clear ferroelectric at ± 2 V. (f) Above the ferroelectric phase transition, polarization reversal was not apparent, verifying ferroelectric behavior below $T_C$.

## CONCLUSIONS

We demonstrate a human-AI collaborative (HAIC) autonomous synthesis workflow that combines retrieval-augmented LLM hypothesis generation and analysis with an autonomous PLD platform to study RE of complex oxides. In the BTO/graphene system, the HAIC campaign efficiently links growth conditions to graphene damage and, together with targeted in situ Raman and plasma plume measurements, shows that gas chemistry dominates overall degradation while ballistic impacts seed reactive defects; this isolates a low-$pO_2$, low-$T_s$ regime that preserves graphene yet conflicts with optimal BTO growth. Guided by this insight, a two-step $Ar/O_2$ sequence is required and can yield ferroelectric, crystalline BTO while maintaining the graphene interlayer. Beyond the specific application, the value lies in the pattern of work: humans and AI collaboratively set objectives, refine processes and analyses between autonomous batches, and redeploy updated policies to achieve rapid and efficient scientific progress. The same HAIC approach transfers to other autonomous platforms and will strengthen with the fast-paced improvement of language models, enabling more reliable and data-efficient autonomous experimentation.

**CONFLICT OF INTEREST**

The authors declare no competing interests.

**AUTHOR CONTRIBUTIONS**

**A.H.**: Conceptualization (equal), Investigation (equal); Writing-Original Draft (equal); Writing-Review & Editing (equal). **D.T.Y.**: Investigation (supporting), Writing-Original Draft (supporting). **J.C.**: Investigation (supporting); Software (supporting). **R.B.**: Investigation (supporting). **I.V.**: Investigation (supporting). **J.L.**: Investigation (supporting). **S.G.**: Investigation (supporting). **C.M.R.**: Investigation (supporting). **K.X.**: Writing-Original Draft (supporting). **Y.L.**: Investigation (supporting). **E.Z.**: Investigation (supporting). Writing-Original Draft (supporting). **R.K.V.**: Writing-Original Draft (supporting); **S.B.H.**: Conceptualization (equal); Investigation (equal); Methodology (lead); Software (lead); Writing-Original Draft (equal); Writing-Review & Editing (lead). All authors read and approved of the final manuscript.

**DATA AVAILABILITY**

The data that support the findings of this study are available upon reasonable request.

**CODE AVAILABILITY**

The code used for Bayesian optimization and surrogate models in this study are openly available through the ALchemy package at https://github.com/sumner-harris/alchemy.git and the user interface software for RAG using the OpenAI API for ChatGPT are available at https://github.com/sumner-harris/RAGAgent_AutonomousMaterialsSynthesis.git

# METHODS

## Retrieval Augmented Generation

Text from 345 articles and supplemental information files were extracted, filtered, split into 300-token chunks with a 30-token overlap, and embedded as 3072-dimensional feature vectors using OpenAI's text-embedding-3-large model. The user query was embedded with the same model and maximal marginal relevance (MMR) search is used to retrieve the top 50 most similar text chunks, balancing high similarity to the query with diversity of sources. These chunks were appended to the initial query to form the full LLM prompt. The LLM used was GPT-5 Thinking. We developed an open-source, user-friendly graphical interface package that constructs a knowledge base from a collection of PDFs and performs RAG with ChatGPT (requires an OpenAI API key). The package can be downloaded from https://github.com/sumner-harris/RAGAgent_AutonomousMaterialsSynthesis.git

## Bayesian Optimization

Bayesian optimization is carried out using a custom python package (ALchemy) built using BoTorch[33] with a single task Gaussian process regression model implemented via GPyTorch[34]. Input features $x$ are normalized [0,1] per dimension and the target was standardized to zero mean and unit variance for training. The GP prior used a learned constant mean function and a radial bias function (RBF) kernel with automatic relevance determination (ARD) wrapped in a scale kernel:

$$k(x,x') = \sigma_f^2 \exp\left(-\frac{1}{2}\sum_{i=1}^{d}\frac{(x_i - x'_i)^2}{\ell_i^2}\right)$$

where $\sigma_f$ is the target variance and $\ell_i$ is the length scale of each dimension. GP hyperparameters (mean constant, target scale, length scales, and noise) were learned by maximizing the marginal likelihood on the transformed data. The autonomous synthesis campaign was posed as a maximization problem and used the Monte Carlo formulation of the log expected improvement acquisition function[35] (qLogEI), implemented via BOtorch. ALchemy is freely available at https://github.com/sumner-harris/alchemy.git

**CVD Growth of Graphene on Cu**

Monolayer graphene was synthesized by atmospheric-pressure chemical vapor deposition (AP-CVD) on copper foil following the ORNL recipe. In brief, 25-125 μm Cu foils were solvent-cleaned (acetone, isopropanol) and electropolished in 85 % $H_3PO_4$ at 1.9 V for 30-60 min, then rinsed with DI water and IPA and dried. Foils were loaded into a horizontal quartz-tube furnace and pre-annealed at 1000 °C for 1 h in a forming-gas carrier (2.5 % $H_2$ in Ar; total flow ~ 5 L $min^{-1}$). Growth proceeded at 1000 °C by stepwise $CH_4$ dosing in the $H_2$/Ar carrier to suppress multilayers: $CH_4$ (1000 ppm in Ar) was metered to give 30, 50, 70, and 100 ppm at the reactor inlet, 45 min per step (total ~3 h) while maintaining the same total flow. At the end of growth, the furnace was opened to accelerate cooling under $H_2$/Ar. This is adapted from Vlassiouk et al. [36], which details the role of Cu pretreatment, staged methane dosing, and lamination-based transfer for large-area monolayer graphene.

**Wet Transfer of Graphene**

The as-grown graphene on the copper foil was spin-coated with PMMA at a speed of 2000 revolutions per minute (rpm) and left to cure at room temperature for 24 hours. An ammonium persulphate solution was employed to etch the copper foil. The remaining graphene, along with the PMMA layer, was transferred to HCl and then to DI water to dissolve any salt residue. The cleaned graphene was subsequently collected and dried on the desired substrate. The PMMA layer was removed by immersing it in acetone and then rinsing it with fresh acetone and isopropyl alcohol to eliminate any PMMA residue.

**Pulsed Laser Deposition**

*Film growth*: BTO films were deposited by PLD using a KrF excimer laser (248 nm, 25 ns, 5 Hz; Coherent LPX 305F) to ablate a stoichiometric BTO target. The target-substrate distance was 5.0 cm, and the substrate temperature was varied from 300-800 °C. Laser fluence at the target was set with a motorized attenuator and bidirectional slits on a projection beamline, defining a spot on the target (area 0.022 $cm^2$) into which all the laser energy that was directed into the deposition chamber (44 mJ in the case of 2 $J/cm^2$) was deposited - i.e., a top-hat beam profile scenario. Background gas during deposition were either Ar (99.999%, 5 sccm) or $O_2$ (99.999%, 5 sccm). Gas flow values specified in the text were established with mass-flow controllers and the total pressure was maintained with a throttle valve. The base pressure prior to each set of ten depositions was < 5 × $10^{-6}$ Torr. Substrates were bonded to Inconel carrier plates that were heated from the backside with a 976 nm, 140 W diode laser, and temperature was controlled in a closed loop fashion with a backside pyrometer with an accuracy of ±1 °C. Font-side temperature (or approximately the substrate temperature) was independently confirmed with either a spot-welded fine gauge thermocouple and Raman shift vs temperature measurements on Si.

*Plume diagnostics.* Ion-probe waveforms were measured with a biased wire probe (-40 V bias; 4 mm length, 0.4 mm diameter) placed adjacent to the substrate with a 4.5 cm target-probe distance. The probe current was recorded on a 1 GHz oscilloscope (Tektronix MSO64) across a 50 Ω feed-through BNC terminator. The oscilloscope was triggered by a fast photodiode monitoring the KrF laser pulses.

**In situ Raman Spectroscopy**

A 532 nm continuous wave laser (Cobolt Samba, 1 W max, Huber Photonics) was used to excite the samples at 55° angle of incidence. The laser spot on the samples was elliptical (0.7 mm × 0.6 mm) with the major axis aligned with the spectrometer slit. Raman scattered light was collected at normal incidence and imaged onto the spectrometer slit with a 75 mm diameter, $f$ = 350 mm spherical lens mounted inside the chamber. A long-pass edge filter (Semrock, RazorEdge) was used to filter the laser line. The spectrometer (Princeton Instruments, Isoplane SCT 320) was coupled with a CCD detector (Princeton Instruments, PIXIS 400BRX) and used a 2400 grooves per mm holographic grating. The laser power and CCD exposure times for all samples and reference spectra were 300 mW and 15 s exposure, 4 averages.

**In Situ Reflectivity Optical Contrast Modeling**

Reflectivity was monitored *in situ* using a stabilized, randomly polarized HeNe laser (632.8 nm, 1.2 mW; HRS015B, Thorlabs, Inc.) incident at 32.5°. A liquid crystal polymer depolarizer (DPP25-B, Thorlabs, Inc.) was used to ensure random polarization. The reflected signal was filtered through a laser line filter (FL632.8-1, Thorlabs, Inc.) and detected using a silicon photodiode (SM1PD1B, Thorlabs, Inc.) connected to a source measure unit (2450, Keithley Instruments). Data were recorded continuously, with a 10 second pre-deposition baseline used to determine the average initial signal, $I_0$. Real-time optical contrast was calculated as $C(t) = (I(t) - I_0)/I_0$ where $I(t)$ is the photodiode current at time t. Deposition was terminated automatically when $C(t)$ reached the target contrast value corresponding to 5 nm of BTO.

The optical contrast of the Vacuum/BTO/graphene/SiO$_2$/Si multilayer stack was modeled using the transfer matrix method (TMM), which calculates the Fresnel reflection coefficients recursively across stratified media. The total reflectance was computed for both s- and p-polarized light and averaged to represent randomly polarized illumination. Refractive indices used at a wavelength of 632.992 nm were: Si (3.872 + 0.016i), SiO$_2$ (1.4689 and 270 nm thickness), monolayer graphene (2.6 + 1.3i), and BTO (2.38 + 0.01i). The BTO film thickness for the BO optimized films were kept for 5.25 nm, corresponding to ~13-unit cells based on a c-axis lattice parameter of 0.404 nm (**Figure S10**).

The calculated reflectance contrast for a 5.25 nm-thick BTO film on this substrate configuration was -0.203, which was used as the target threshold for terminating growth in real-time during deposition. Temperature-dependent variations in refractive index were neglected. Simulations showed that angular uncertainty of ±2.5° from the nominal incident angle of 32.5° resulted in negligible variation in the calculated contrast, remaining below the experimental noise floor.

**Ex situ Characterization of Thin Films**

XRD and XRR were done with a PANalytical X'pert PRO MRD diffractometer implementing a hybrid monochromator (Cu K$\alpha_1$) with a 1/8° and 1/32° divergence slit, respectively. A 0.27° parallel plate collimator with 0.04 radian Soller slits were used on the detector (Xe proportional counter) side in both cases and an additional 100 μm receiving slit for XRR.

PFM was performed in an Oxford Instruments Asylum Research Vero with an Environmental Scanner using a BudgetSensors ElectriMulti75-G Pt-Si tip. Measurement was performed off-resonance with a drive amplitude of 500 mV. To gather temperature-dependent SS-PFM data, the sample was heated to 250° C, which was verified with a thermocouple. Data was collected off-bias at 0.1 Hz and with a 2-V drive oscillation. 3 consecutive loops were taken at five random locations; the presented data is the average of 15 separate loops. The sample was then cooled below 100° C and measurements were repeated. SS-PFM data was plotted using pre-built functions in the *Hystorian* package[37].

**Molecular Dynamics Simulations**

The MD simulations were performed using the LAMMPS[38] MD simulation package. For the impact simulations, the AIREBO[39] potential was used to describe graphene, and the Tersoff[40] potential was employed for the amorphous SiO$_2$ substrate, while the interactions between the graphene monolayer and the substrate were described using Lennard-Jones[41] (LJ) parameters, similar to Zhao and Xue[42]. Short-range repulsive Ziegler-Biersack-Littmark[43] (ZBL) potentials were used in all pair interactions, including the interactions with the projectile, Ba. The size of the systems used for the impact simulations was about 10.1 nm × 10.2 nm. The substrate size was 5 nm, and the graphene monolayer contained about 4000 atoms. The systems were relaxed and then equilibrated at 600 °C (873.15 K). Projectile energies of 139 eV/Ba-atom and 318 eV/Ba-atom (approximately the maximum observed) were used to study the ballistic defect formation. A total of 200 Ba ions and a total of 133 Ba ions were used for each energy, respectively. The impact of the ions with the system was simulated in the microcanonical ensemble (NVE), with a 5 ps interval between successive ions, which was sufficient for system thermalization. A variable timestep was used to account for the impact dynamics, with a maximum value of 0.5 fs. Periodic boundary conditions were applied along the x and y dimensions of the MD box.

For the annealing simulations, a ReaxFF force field for graphene and oxygen interactions[44] was utilized in a triclinic simulation cell was used with periodic boundary conditions applied along the x and y directions and a free surface along z in vacuum, with cell dimensions of about 49.3 Å (x-axis) and 42.7347 Å (y-axis). The system consisted of a graphene bilayer, where the bottom monolayer was thermostatted to mimic the thermal response of a substrate. The top monolayer was left unrestrained and interacted directly with the incoming gas species. In the figures only the upper monolayer is shown for simplicity, since the lower monolayer is thermostatted and does not interact with the gas. The annealing simulations were performed at 0 °C (273 K), 300 °C (573 K), 600 °C (873 K), 700 °C (973 K), and 800 °C (1073 K), after prior equilibration of each system at the corresponding temperature. Annealing simulations were performed under the NVE ensemble using a variable timestep (maximum 0.25 fs) to capture fast dynamics that may occur. To simulate the gas, we inserted 1000 O atoms with intervals of 125 fs. The total simulation time at each temperature was about 1 ns.

**Exfoliation of BaTiO$_3$**

To exfoliate the BTO film, a 2.7 μm thick Ni stressor layer was deposited on the two-step-grown BTO/Gr/STO heterostructure using DC magnetron sputtering at 30 W, under a base pressure of 1x10$^{-6}$ Torr. The resulting Ni/BTO stack was subsequently exfoliated using a thermal release tape (TRT). The inset of **Figure 4b** displays the exfoliated BTO/Ni layer adhered to the TRT.

## ACKNOWLEDGEMENTS

This work was supported by the Center for Nanophase Materials Sciences (CNMS), which is a US Department of Energy, Office of Science User Facility at Oak Ridge National Laboratory. Materials synthesis with *in situ* diagnostics was supported by the U.S. Department of Energy, Office of Science, Basic Energy Sciences, Materials Sciences and Engineering Division. LLM interface development was supported by the US Department of Energy, Office of Science, Office of Basic Energy Sciences, MLExchange Project, award number 107514. PFM was sponsored by the INTERSECT Initiative as part of the Laboratory Directed Research and Development Program of Oak Ridge National Laboratory, managed by UT-Battelle, LLC for the US Department of Energy under contract DE-AC05-00OR22725. This research used resources of the Compute and Data Environment for Science (CADES) at the Oak Ridge National Laboratory, which is supported by the Office of Science of the U.S. Department of Energy under Contract No. DE-AC05-00OR22725.

## SUPPORTING INFORMATION

XRD of BTO/Gr/SiO$_2$/Si (two-step); surrogate (BNN/RF) models corroborating GP response surfaces; verbatim LLM-generated hypothesis and autonomous-PLD plan; human-AI "offline conferences" refining Raman preprocessing/curve-fitting (shared normalization, local baselines, contaminate peaks, D′ inclusion, global gain correction). Ion-probe TOF analysis and GP mapping of plume MPKE vs pO$_2$, fluence. MD simulation of Ba ion bombardment of graphene on an SiO$_2$ substrate with different Ba number of ions and energy. Raman heat maps with pulses and ambient; annealing Raman (no D-band activation); ferroelectric BTO characterization (PFM/XRR); reflectance-contrast vs BTO unit cells. Comparison with literature for PLD-based remote epitaxy and exfoliation summary.

# Supplementary Information for:

# Human-AI collaborative autonomous synthesis with pulsed laser deposition for remote epitaxy

*Asraful Haque[1], Daniel T. Yimam[1], Jawad Chowdhury[1], Ralph Bulanadi[1], Ivan Vlassiouk[1], John Lasseter[1], Sujoy Ghosh[1], Christopher M. Rouleau[1], Kai Xiao[1], Yongtao Liu[1], Eva Zarkadoula[1], Rama K. Vasudevan[1], Sumner B. Harris[1]\**

2. Center for Nanophase Materials Sciences, Oak Ridge National Laboratory, Oak Ridge, Tennessee 37831, United States.

   *Correspondence should be addressed to: harrissb@ornl.gov

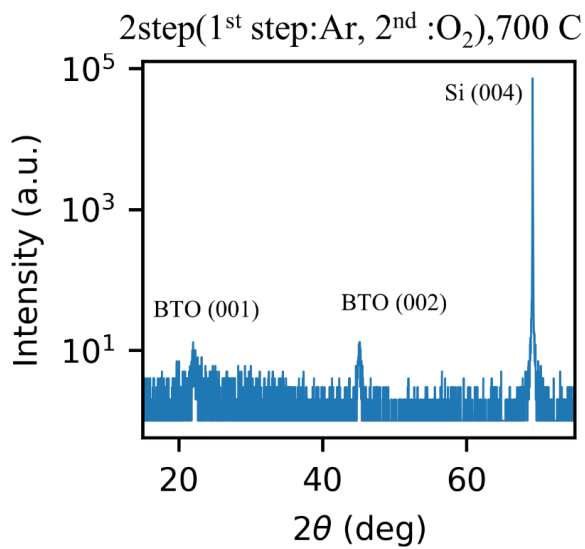

**Figure S1**. XRD θ-2θ scan of BTO grown on Gr/SiO$_2$/Si using two step deposition shows c-axis orientation.

**Note S1 – Verbatim LLM Hypothesis**

Using the RAG pipeline with our knowledge base, we "chatted" with the LLM to gain background on remote epitaxy, PLD of BTO, Raman analysis of graphene, providing details of our experimental capabilities. After these chats, we prompted the LLM to generate a hypothesis and propose and experiment, which is the hypothesis/experiment that we chose as a basis for the HAIC autonomous workflow presented in the main text.

<u>User prompt:</u>

> Based on the previous conversation, given below, state a testable hypothesis for filling the knowledge gaps in remote epitaxy of PLD using autonomous synthesis with in situ Raman spectroscopy. Suggest an experiment and provide the parameter ranges. Use only 2 to 4 PLD parameters. Target-to-substrate distance is not a parameter that we can tune so do not suggest that one.
>
> --- Previous Conversation ---
>
> [All previous prompts and responses were appended here]

<u>ChatGPT said:</u>

> There exists a well‐defined "soft‐landing" window of PLD growth parameters—specifically O$_2$ background pressure (pO$_2$), laser fluence (F), and substrate temperature (T$_s$)—in which (a) the kinetic energy of both plume species and sputtering‐gas neutrals falls below the displacement threshold of graphene (<~22 eV/C‐atom) and (b) BaTiO$_3$ still nucleates in exact registry to the underlying polar substrate through the graphene interlayer. In that window, the Raman defect metric $I(D)/I(G)$ will remain below 0.1 (indicating negligible graphene damage), and ex situ structural probes (RHEED, XRD pole figures) will confirm remote epitaxy (single‑crystal, substrate‑matched $BaTiO_3$). Outside this window, either graphene damage rises sharply ($I(D)/I(G) \gg 0.1$) or film registry is lost (transition to van der Waals epitaxy).

Proposed Autonomous‑PLD Experiment

- Knobs (3):
  - $pO_2$: 50 to 250 mTorr (steps of ~50 mTorr)
  - F (KrF excimer, 248 nm): 1.2 to 2.2 J/cm² (steps of ~0.2 J/cm²)
  - $T_s$: 650 to 800 °C (steps of ~50 °C)

- Fixed parameters:
  - Repetition rate: 5 Hz
  - On-axis geometry, monolayer graphene on $TiO_2$-terminated $SrTiO_3$(001)
  - Deposition thickness per run: ~10 nm of $BaTiO_3$

- Workflow:

AI (Bayesian‑optimization) proposes ($pO_2$, F, $T_s$).

Deposit ~10 nm $BaTiO_3$ by PLD.

Immediately acquire in situ Raman on graphene to extract $I(D)/I(G)$.

Feed $I(D)/I(G)$ into the optimizer (target: minimize $I(D)/I(G)$).

Iterate for ~20–30 runs until convergence of parameters.

Grow a full‑thickness film at the optimized conditions and confirm remote epitaxy by RHEED oscillations and XRD pole‑figure analysis.

Expected Outcome

- A well‑localized optimum, e.g. $pO_2 \approx$ 120–160 mTorr, F $\approx$ 1.6–1.8 J/cm², $T_s \approx$ 700 °C, yielding $I(D)/I(G) <$ 0.1 (minimal graphene damage) and single‑crystal $BaTiO_3$ registry.
- Outside this window: at lower $pO_2$ (<100 mTorr) or higher F (>2.0 J/cm²), fast neutrals and high‑energy species induce graphene defects ($I(D)/I(G) \gg 0.1$) [Chen et al. APL 103,033109 (2013) | chunk 14]; at higher $T_s$ (>800 °C) graphene oxidizes [Wohlgemuth et al. APL Mater.12,021113 (2024) | chunk 5], and at too-low $T_s$ (<650 °C) $BaTiO_3$ fails to crystallize [Choi et al. Science 306,1005 (2004) | chunk 8].

Filling the Gap

This experiment directly correlates PLD variables to graphene integrity (via in situ Raman) and epitaxial registry, establishing quantitative growth rules for damage‑free remote epitaxy by PLD—a gap in the current mechanistic understanding.

**Bayesian Neural Network**

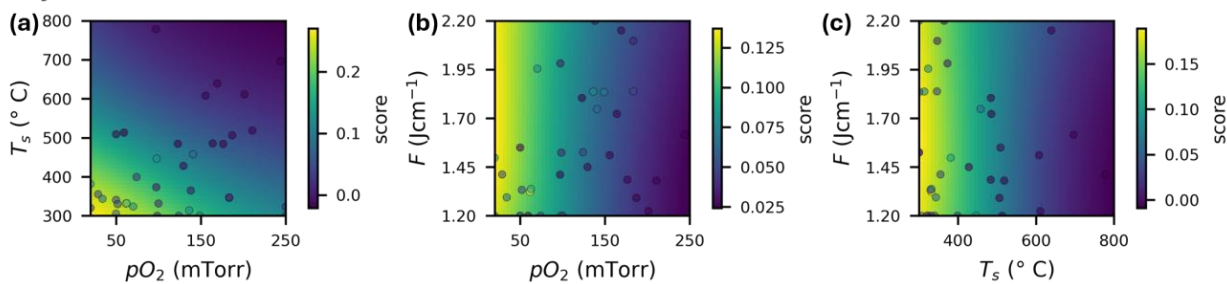

**Random Forest**

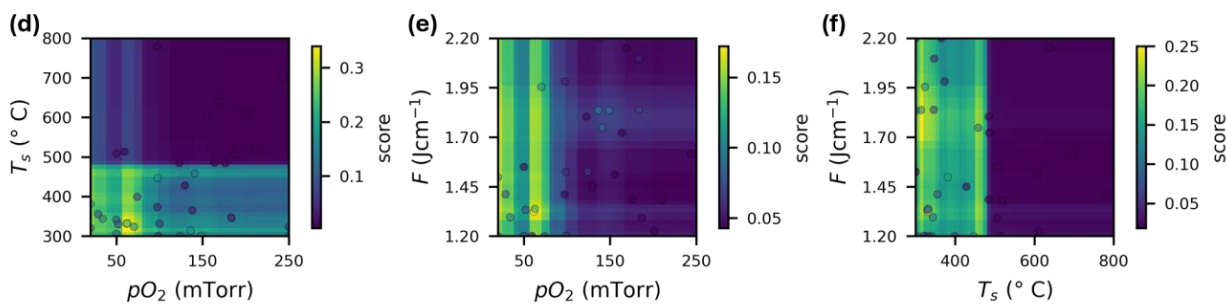

**Figure S2**. Kernel-free surrogate models fit with the same data from **Figure 2a-c** in the main text. (a-c) Bayesian neural network (BNN) and (d-f) Random forest (RF) surrogate models yield similar response surfaces to the Gaussian process (GP) surrogate which aid in validating the GP results. The BNN predictions are nearly identical to the GP predictions while the RF displays the usual "blocky" lookup-table-like surface. All 3 surrogate models have similar maxima regions and qualitative shapes.

**Note S2 – Notable aspects of offline human-AI conference after phase 1**

After phase 1 of the autonomous BTO growth (10 samples grown), the offline human AI conference resulted in the first of several changes to the Raman analysis and found an outlier point (caused by bad curve fitting) which skewed the GP. Below, we show an excerpt from this human-AI conference to highlight the automatic data analysis the LLM conducted to identify the outlier and the recommended fixes. Parts of the prompts and responses are abbreviated in brackets for brevity.

User prompt:

> *[PNG images of all 10 Raman spectra fits were uploaded to the LLM]*
>
> I am fitting Raman spectra of graphene before and after depositing BaTiO3 on it to search for conditions which lead to the maximum of our score metric. My score should be bounded between 0-1 but the fitting function is not working robustly, leading to bad fits and anomalously large Y values, which may mislead my gaussian process surrogate model during Bayesian optimization. Please analyze the current graphene fitting function code and plot results so far as well as the X,Y values used in the BO. X values are [Pressure O2, Temperature, Laser energy] and Y is the score returned from the graphene_objective function. Several of the samples show that the graphene is completely destroyed after deposition which is expected. But it is clear that the fitting function and score calculation is returning unexpected values. Plots of the graphene fitting results are attached.
>
> --- Current X, Y Data for GP model ---
>
> X=(tensor([[ 61.6309, 331.4551, 29.1189], [176.3493, 484.1429, 30.4635], [201.3447, 611.2147, 26.9216], [210.6028, 518.4784, 30.3651], [ 97.0653, 778.7671, 31.0331], [169.2077, 639.2903, 47.3119], [187.0194, 506.3742, 28.4088], [155.4756, 607.8483, 33.2359], [243.8909, 696.5065, 35.5431], [ 59.3433, 513.2381, 26.4000]], dtype=torch.float64), Y=tensor([[5.4303e-01], [4.4023e-08], [7.7948e-01], [1.6241e-22], [4.0972e-01], [2.3269e+04], [5.7596e-02], [5.3415e-08], [1.1697e-14], [1.8283e-02]], dtype=torch.float64))
>
> --- Code for graphene fitting ---
>
> *[The full code for the graphene fitting functions is pasted here]*

ChatGPT said:

> *[ChatGPT created scatter plots of the score vs growth parameter, shown below is just 1 of the 3 plots to display the outlier point]*

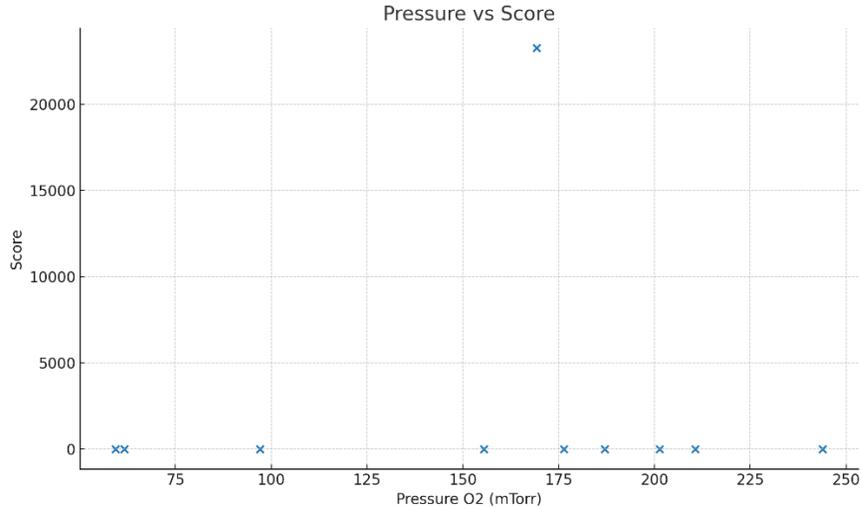

We can see straight away from the table and scatter‑plots that, while most of your scores lie nicely in [0,1], the point at (P≈169 mTorr, T≈639 °C, E≈47 mJ) shoots up to ≈2.3 × $10^4$. That single outlier will dominate your GP and lead BO astray.

What's going wrong in your fitting / scoring

1. Normalization by a tiny "global max".

When the graphene is effectively destroyed, your filtered intensity has almost no G-peak and the global maximum (used to normalize) collapses to some small noise level. Dividing by this tiny number wildly amplifies both D and G areas and produces huge ratios.

2. Dividing by near-zero denominators.

You compute

$rg = I_{G,treated}/I_{G,ref}$, $\Delta = I_{D,treated}/I_{G,treated} - I_{D,ref}/I_{G,ref}$, $score = rg/(1+\Delta)$.

If $I_{G,ref}$ or $1+\Delta$ is nearly zero you get a numerical blow-up.

3. Poor backgrounds → bad peak‑area estimates.

A simple quadratic background fit over the entire 1100–1700 $cm^{-1}$ range is not robust when only a tiny residual "bump" remains.

Recommended mitigations

1. Reject or clamp destroyed samples up front.

If the total G-peak amplitude from the reference is below some threshold (e.g. $I_{G,ref} < \epsilon$), or if the treated spectrum shows no discernible G, assign a safe score (e.g. zero) rather than trying to fit.

2. Normalize to a more stable reference.

Instead of global_max of the treated spectrum, normalize both reference and treated to the same baseline. For example, fit the G-peak height on the reference and divide both spectra by that. That way a missing peak on the treated side gives rg≈0 rather than ∞.

3. Bound your score.

After computing, clip: score = max(0.0, min(score, 1.0))

This prevents any outliers from wrecking your surrogate's kernel fit.

4. Improve background fitting locally.

Fit the baseline only in flanking regions around each peak (e.g. [1500–1550] and [1650–1700]) rather than across the entire Raman window. This gives a more accurate area under the peak and avoids large negative residuals.

5. Add fit quality checks.

Inspect result_trt, redchi or the ratio of residual RMS to peak height. If the fit quality is poor (e.g. reduced $\chi^2 \gg 1$), again fallback to a default score.

With those changes you'll keep all returned scores safely in [0,1], and the BO surrogate will see a consistent, meaningful response surface rather than be dominated by a single numerical artifact.

Following this initial suggestion, we continued a lengthy exchange with the LLM to update and test the fitting functions, refining the LLM's initial suggestions.

**Note S3 - LLM assisted iterative refinement of Raman analysis**

The LLM assisted in 3 revisions of the graphene fitting function after phase 1, phase 2, and phase 3 of the human-AI cooperative autonomous synthesis experiment. Revision 1 introduced a shared normalization, peak flank-based local baseline, and a 3-peak model which conditionally added polymer contaminate peaks. Including contaminate peaks remedied a "known unknown" in the spectra that we did not realize we needed to account for before experiments. Revision 2 added the D' peak to the fit and a G peak amplitude-based retention ratio (changing from an area ratio) to improve fitting in the G region. Finally, revision 3 added a global gain correction to re-scale the "after deposition" spectra to account for the increased background scattering caused by the BTO deposition (an example of dealing with an "unknown unknown" effect that was only found after conducting experiments.) Below, the changes are detailed in **Table S1**.

**Table S1.** Summary of Raman curve fitting function updates resulting from offline human-AI conferences

| Aspect | Original | rev1 | rev2 | rev3 |
|---|---|---|---|---|
| Pre-baseline step | none | Subtract last point | same | same |
| Shared normalization | No (each by own max) | Yes (norm_ref from ref) | Yes | Yes |
| Local baseline (flanks) | In-model poly bkg on full region | Yes (1100–1300 & 1650–1700) | Yes | Yes (with data window effectively 1200–1700) |
| Data window for fit (cm$^{-1}$) | 1100–1700 | 1100–1700 | 1100–1700 | 1200–1700 (code) |
| Default peaks | D, G | D, 1450, G | D, 1450, G, D' | D, 1450, G, D' (+ initial guesses) |
| D' > G constraint | — | — | Yes | No |
| Contaminants (1480, 1520) | — | Conditional (only if fit poor) | Always (if-check disabled) | Always |
| Gain correction (treated↔ref) | — | — | — | Yes (s_est) |
| r_g definition | G area ratio | G area ratio | G amplitude ratio | G amplitude ratio (after gain correction) |
| Reference gate (I_g_ref, redchi) | — | Yes | Yes | Removed (only norm_ref too small aborts) |

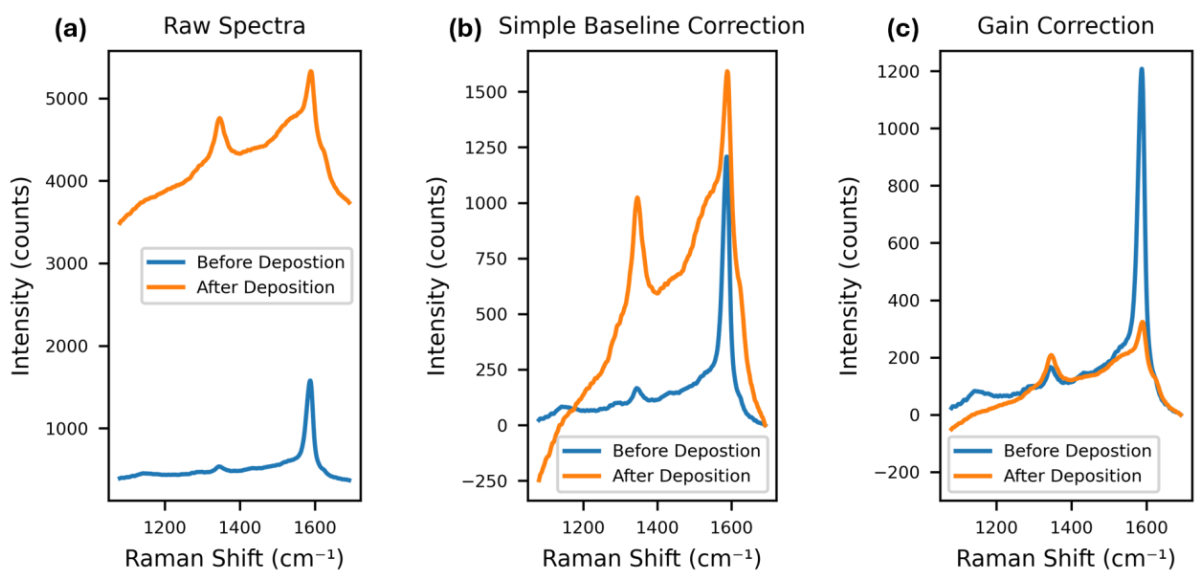

**Figure S3**. Representative example of the final preprocessing procedure for Raman spectra before curve fitting for the sample grown at $pO_2$ = 50 mTorr, $T_s$ = 305 °C, F = 1.2 J cm$^{-2}$. (a) Raw Raman spectra after deposition have increased background scattering due to the presence of the BaTiO$_3$ thin film, which complicates the relative analysis of Raman peak areas. (b) A simple baseline correction is applied by subtracting the intensity of the last data point. This is not enough because it is physically impossible for the G peak area to increase after deposition (potential for graphene to etch/oxidize implies G/G$_0$ ≤ 1) (c) Finally, the global scale is estimated and the "after deposition" spectrum is corrected to yield the final preprocessed spectra for curve fitting. The global scale ($s$) is estimated as the median of the ratio of the D and G peak region median intensities i.e. $r_D = m_{D,after}/m_{D,before}$ , $r_G = m_{G,after}/m_{G,before}$ , give the global scale factor $s = median(r_D, r_G)$. The "after deposition" spectrum is divided by s to correct the intensity. Notably, this generally maintains the intensity of polymer contaminate peaks, which should not change significantly (see **Figure S4** below). The global scale factor was $s = 4.912$ in this case.

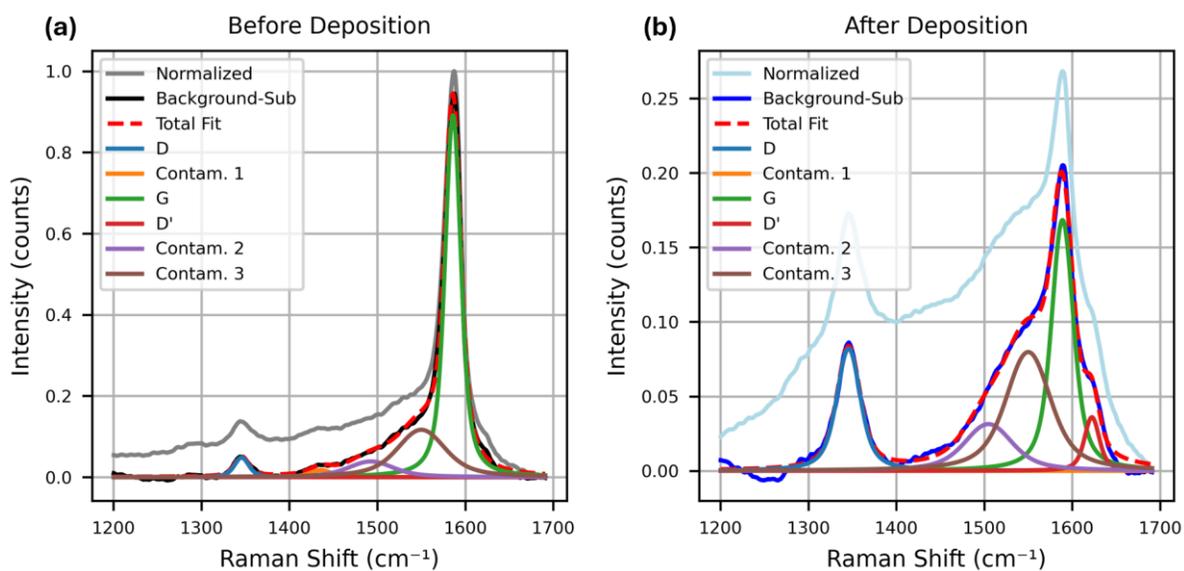

**Figure S4**. Representative example of graphene Raman spectra curve fitting procedure, following the preprocessing for intensity correction shown in Figure S3, for the sample grown at $pO_2 = 50$ mTorr, $T_s = 305$ °C, $F = 1.2$ J cm$^{-2}$. Each spectrum is fit to extract $I_D$ and $I_G$ accounting for polymer contaminate peaks and a polynomial background. (a) Before BaTiO$_3$ deposition and (b) after BaTiO$_3$ deposition. Note that the polymer contaminates peaks (Contam. 2, Contam. 3) approximately maintain their absolute intensity, which helps verify that the global intensity correction applied to the "after deposition" spectrum is appropriate. For this sample, the change in $I_D/I_G$ was $\Delta = (0.05-0.529) = 0.480$ and the retention ratio was $r_G = 0.259$ to give a score of 0.175.

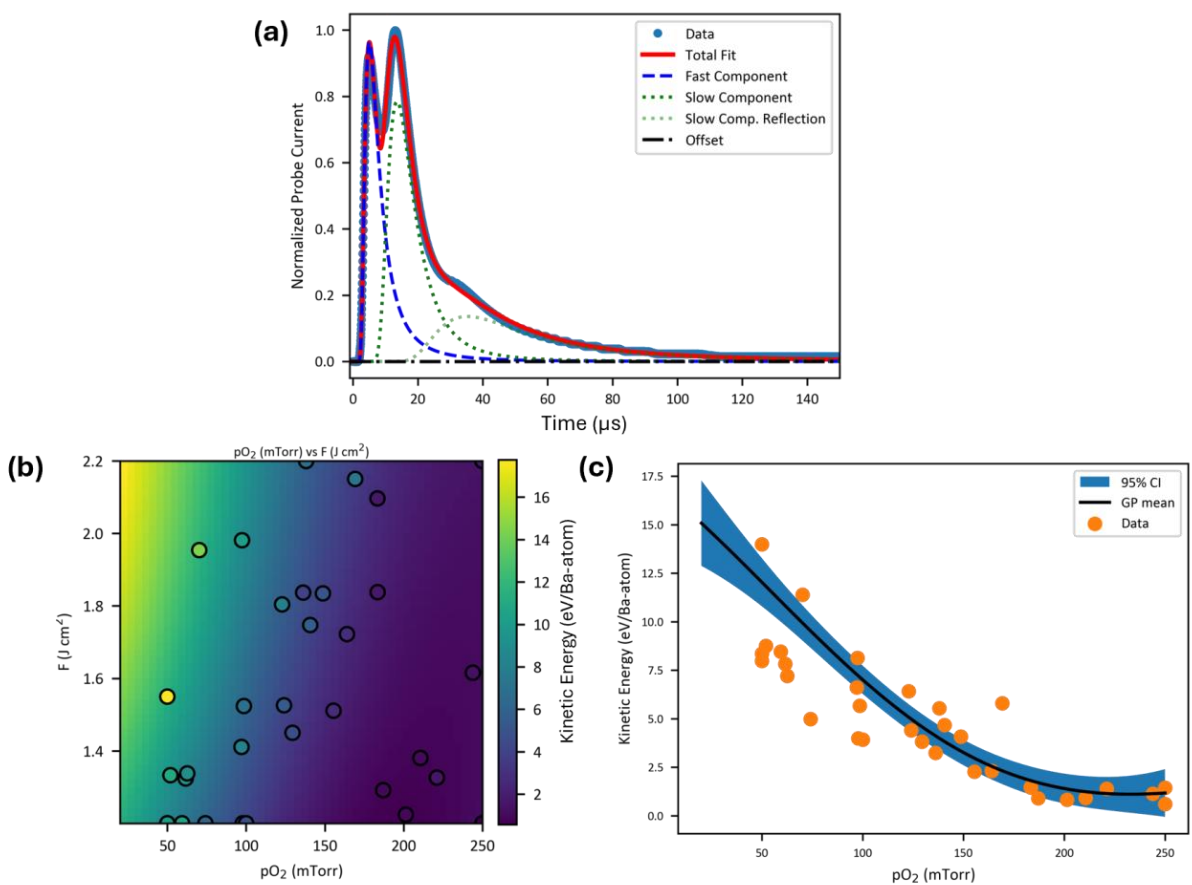

**Figure S5**. Most probable kinetic energy (MPKE) of the slow component of the BaTiO$_3$ plume was fit from the ion probe collected during the autonomous experiment. (a) Example ion probe trace taken at pO$_2$ = 61.6 mTorr and F = 1.32 J cm$^{-2}$ showing the "fast" component, the scattered "slow" component, and an artifact peak (slow comp. reflection) arising from the plume "bouncing off" the substrate holder and interacting with the ion probe again. MPKE is taken from the time-of-flight (TOF) of a component peak position and calculated as $MPKE = 0.5\, m\, (D/TOF)^2$ where D is the target-probe distance which is 4.5 cm in this case and $m$ is the atomic mass of Ba (137.33 u). (b) Gaussian process model mean predictions of the slow component in the F, pO$_2$ parameter space. (c) GP predictions averaged over fluence displays the usual exponential decay behavior of

**(a)** 139 eV Ba - After 200 ions  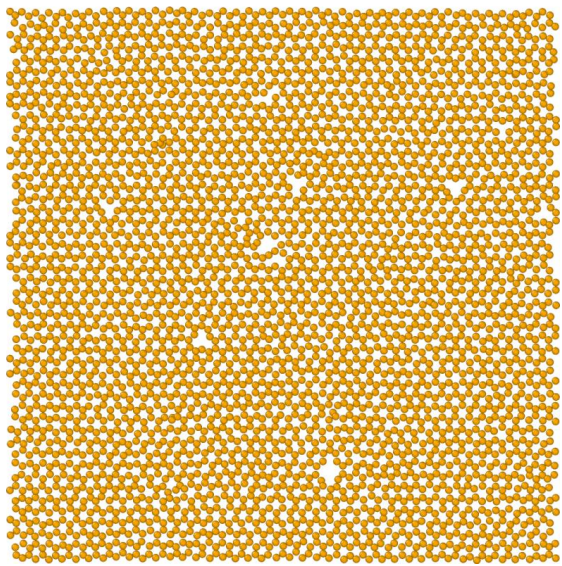

**(b)** 318 eV Ba After 133 Ba ions  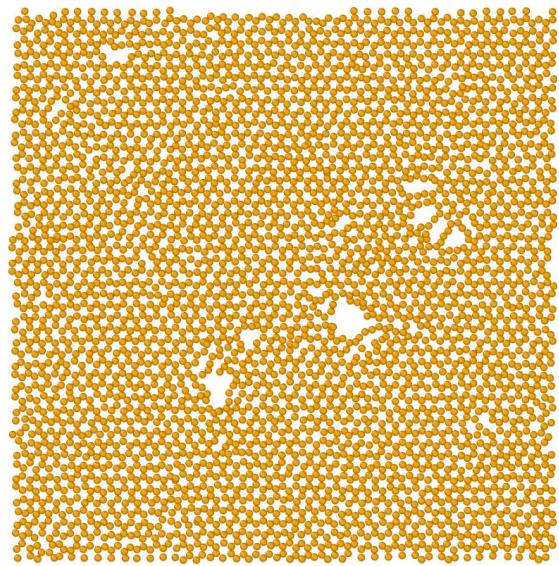

**Figure S6.** Final results of Ba ion bombardment of graphene on an SiO$_2$ substrate. (a) After 200 ions with kinetic energy of 138 eV/Ba-atom. (b) After 133 ions with kinetic energy of 318 eV/Ba-atom.

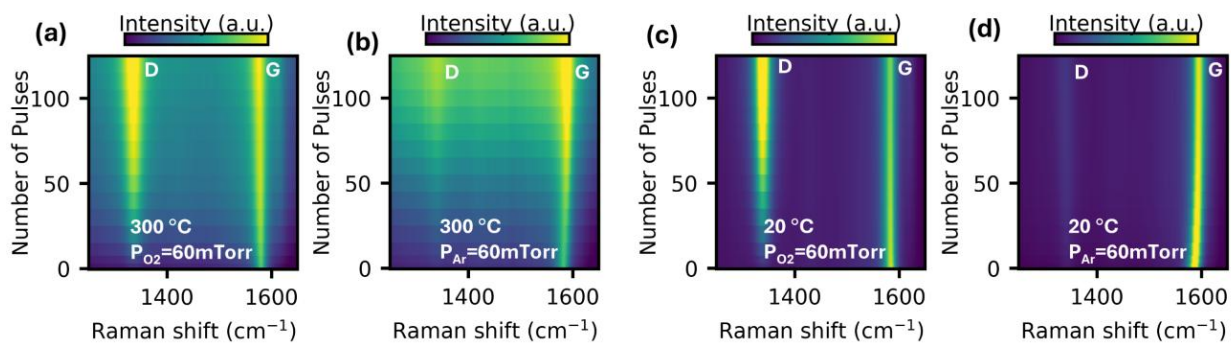

**Figure S7**. Raman heat maps at 300 °C versus laser pulses in (a) Ar and (b) $O_2$. Raman heat maps at 20 °C versus laser pulses in (c) $O_2$ and (d) Ar.

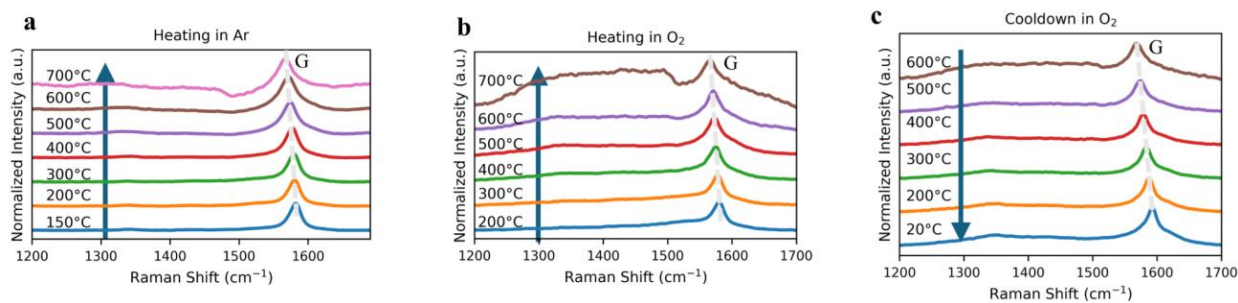

**Figure S8**. Raman spectra of Graphene on $SiO_2$/Si collected during annealing: (a) Heating in Ar (150 °C to 700 °C): no prominent D band; (b) Heating in $O_2$ (200 °C to 700 °C): similar absence of D activation; (c) Cooldown in $O_2$ (600 °C to 20 °C): G shifts without D-band activation. We observe G shifts after cooldown indicating charge transfer rather than disorder, marking the usual p-type doping when transferred graphene is annealed on SiO2.

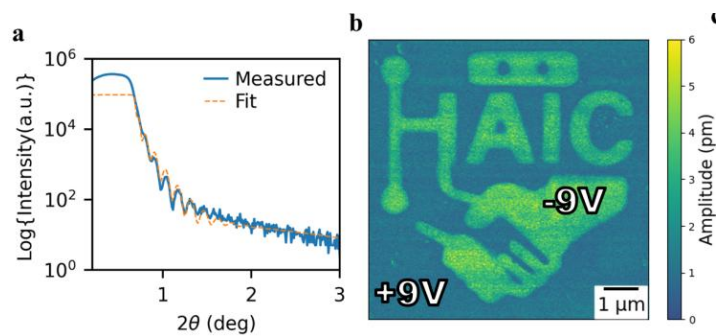

**Figure S9.** (a) X-ray reflectivity (XRR) of BTO (~50 nm) grown on Nb-STO substrate following two step growth method. (b) PFM amplitude of 2-step grown BTO on Nb-STO substrate after patterning.

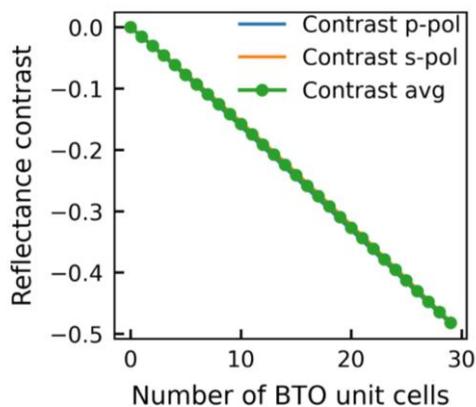

**Figure S10.** Reflectance contrast vs number of BTO unit cell for a heterostructure of BTO on monolayer graphene coated $SiO_2$/Si substrate.

**Table S2**: PLD grown oxides using remote epitaxy and exfoliation status: data extracted using RAG tool.

| S.No | Heterostructure | Synthesis method | Graphene (synthesis method and no. of layers used) | Exfoliation demonstrated | Ref |
|---|---|---|---|---|---|
| 1 | BTO/Gr//STO | PLD | SiC, Bilayer | Yes | 1 |
| 2 | BTO/Gr//STO | MBE | SiC, Bilayer | Yes | 2 |
| 3 | BTO/Gr//STO | PLD | CVD, Bilayer | Yes | 3 |
| 4 | BTO/Gr/SVO//STO | PLD | CVD, Mono | Transferred utilizing graphene enhanced dissolution of SVO | 4 |
| 4 | $VO_2$/Gr//Sapphire | PLD | CVD, Monolayer | < 10 μm | 5 |
| 5 | STO/Gr//STO | MBE | CVD, Monolayer | Yes | 6 |
|   | BTO/Gr//Ge | PLD | Graphene grown on Ge substrate CVD | yes | 7 |
| 6 | STO/Gr//STO | PLD | CVD, Monolayer | No | 8 |
| 7 | $HfO_2$/Gr//STO | PLD | CVD, Monolayer | No | 9 |
| 8 | BTO/Gr//STO | PLD | CVD, Monolayer | Yes | This work |